# Quasi 2-D magnetism in the Kagome layer compound FeSn

Brian C. Sales, Jiaqiang Yan, William R. Meier, Andrew D. Christianson, Satoshi Okamoto and Michael A. McGuire

*Materials Science and Technology Division, Oak Ridge National Laboratory, Oak Ridge TN. 37831*

**This manuscript has been authored in part by UT-Battelle, LLC, under contract DE-AC05-00OR22725 with the US Department of Energy (DOE). The US government retains and the publisher, by accepting the article for publication, acknowledges that the US government retains a nonexclusive, paid-up, irrevocable, worldwide license to publish or reproduce the published form of this manuscript, or allow others to do so, for US government purposes. DOE will provide public access to these results of federally sponsored research in accordance with the DOE Public Access Plan (http://energy.gov/downloads/doe-public-access-plan)."''**

Quasi 2-D magnetism in the Kagome layer compound FeSn

Brian C. Sales, Jiaqiang Yan, William R. Meier, Andrew D. Christianson, Satoshi Okamoto and Michael A. McGuire
*Materials Science and Technology Division, Oak Ridge National Laboratory, Oak Ridge TN. 37831*

**Abstract**

Single crystals of the single Kagome layer compound FeSn are investigated using x-ray and neutron scattering, magnetic susceptibility and magnetization, heat capacity, resistivity, Hall, Seebeck, thermal expansion, thermal conductivity measurements and density functional theory (DFT). FeSn is a planar antiferromagnet below $T_N$ = 365 K and exhibits ferromagnetic magnetic order within each Kagome layer. The in-plane magnetic susceptibility is sensitive to synthesis conditions. Resistivity, Hall and Seebeck results indicate multiple bands near the Fermi energy. The resistivity of FeSn is ≈ 3 times lower for current along the stacking direction than in the plane, suggesting that transport and the bulk electronic structure of FeSn is not quasi 2D. FeSn is an excellent metal with $\rho(300K)/\rho(2K)$ values ≈100 in both directions. While the ordered state is antiferromagnetic, high temperature susceptibility measurements indicate a ferromagnetic Curie-Weiss temperature of 173 K, reflecting the strong in-plane ferromagnetic interactions. DFT calculations show a 3D electronic structure with the Dirac nodal lines along the K-H directions in the magnetic Brillouin zone about 0.3 eV below the Fermi energy, with the Dirac dispersions at the K points gapped by spin-orbit coupling except at the H point. The magnetism, however, is highly 2D with $J_{\text{in-plane}}/J_{\text{out-of-plane}}$ ≈ 10. The predicted spin-wave spectrum is presented.

## I. INTRODUCTION

Magnetic ions arranged in two-dimensional Kagome layers combine interesting topological behavior with a complicated magnetic response due to frustration and competing magnetic interactions [1,2]. For example, the compound $Fe_3Sn_2$ is a canted ferromagnet with $T_c$ = 647 K with interesting topological properties, including a large intrinsic anomalous Hall conductivity [3,4] that has been associated with Weyl nodes near the Fermi energy [3]. There are also reports that the Weyl nodes are switchable in a magnetic field [5] and that $Fe_3Sn_2$ also hosts magnetic skyrmion bubbles at room temperature [6].

The structure of $Fe_3Sn_2$ [7] contains 2D Kagome nets of Fe with the hexagonal holes filled with Sn as shown in Figure 1a. This gives a composition of $Fe_3Sn$ for the layers containing the Kagome nets. Indeed, the related compound $Fe_3Sn$ can be viewed as an infinite stack of such layers fused to one another by interlayer bonding  (Figure 1b). In $Fe_3Sn_2$ fused pairs of $Fe_3Sn$ layers are separated by a layer of Sn (Figure 1c). In FeSn, the topic of the present study, single $Fe_3Sn$ layers are separated by layers of Sn (Figure 1d) [8]. As the Kagome nets of Fe are spatially decoupled, the magnetic order becomes come complicated and the transition temperature is suppressed. $Fe_3Sn$ is a ferromagnet with $T_C = 727$ K, $Fe_3Sn_2$ is a canted antiferromagnet with $T_C = 647$ K, and FeSn exhibits a complex antiferromagnetic structure below $T_N = 365$ K [7].

The similarity of the crystal structures of FeSn and $Fe_3Sn_2$ and the general interest in the topological properties of Kagome layers provided the motivation for the present study.

The crystal structure and magnetism of polycystalline FeSn has been investigated previously using x-rays [7,8] , Mossbauer [9-17], and neutron diffraction [16,18]. There is still some uncertainty as to the exact magnetic structure but it is agreed that FeSn orders in an antiferromagnet structure below $T_N \approx 365$ K with the Fe moments in each Kagome layer aligned more or less ferromagnetically in the basal plane and with the direction of the Fe moments alternating between layers stacked along the $c$ axis (Fig. 2). Well below $T_N$ it is likely that the direction of the Fe moments within the plane change with temperature with the spin direction ranging from the (1,0,0) to the (2,1,0) directions [13,16] The antiferromagnetic order doubles the unit cell along the $c$ direction.

In the present work we report an investigation of the synthesis, electronic structure, magnetic, thermodynamic, and transport properties of FeSn single crystals. We conclude that, although the Neel temperature is always close to $T_N$, the in-plane magnetic structure is sensitive to synthesis conditions, which may account for some of the differences reported in the literature. These results also suggest that the magnetic structure should be sensitive to pressure and chemical doping. Our powder neutron diffraction data indicates ferromagnetic planes of Fe coupled antiferromagnetically along the **c** axis (Fig. 2), in agreement with previous reports [16,18]. High temperature magnetic susceptibility data indicate ferromagnetic interactions and little orbital anisotropy. Resistivity, Hall and Seebeck data show that multiple bands are present near the Fermi energy and from transport measurements the electronic structure is fairly isotropic as opposed to quasi-2D. This is consistent with our and other very recent DFT results [ 19-21]. Our DFT calculations, however, indicate that the magnetism is highly 2D with interesting spin wave modes.

## II. EXPERIMENTAL DETAILS

Single crystals of FeSn are grown out of a Sn flux. Guided by the phase diagram reported by Giefers *et al.*[7], 34 g of Sn and 0.33 g of Fe are loaded into a 10 cc alumina crucible and sealed in a silica ampoule under vacuum. The ampoule is heated to 1100 °C, soaked for 12 h, cooled to 1000 °C and soaked for 48 h with occasional shaking of the ampoule, cooled to 800 C at 6 °C/h, and then cooled to 600 °C at 1°C/h. Near 600 °C the excess Sn flux is centrifuged into another 10 cc crucible filled with quartz wool. For transport or magnetic measurements, the excess Sn flux is either mechanically removed or removed by etching in concentrated HCl until the bubbling stopped. Ground crystals are used for powder x-ray or neutron diffraction experiments.

The crystals tend to grow along the *c* axis as long bars, but there are usually a few blocky hexagonal plates (Fig. 2). The crystals exhibit brittle fracture and weak cleaving perpendicular to the **c** axis. The largest crystals weigh about 0.3 g (Fig. 2). The properties of crystals quenched from temperatures near 600 °C are the major focus of this study, however we note that FeSn crystals can be prepared over a wide final temperature from about 520 °C to 760 °C. We find that while $T_N$ is the same for crystals prepared near 760 °C, the in-plane magnetic structure varies as is evident from magnetic susceptibility measurements with H perpendicular to *c*. Crystals prepared at the higher temperatures display a much larger variation in the in-plane anisotropy, similar to that reported recently in [19]. For example, at 2 K in an applied field of 10 kOe, crystals spun at 682 °C exhibit a 6% variation in $\chi_{ab}$ as the crystals are rotated in the *a-b* plane, while for crystals spun near 600 °C the variation is only about 1%. Energy dispersive x-ray measurements on crystals prepared at different temperatures indicate about 0.5% more Fe in the crystals spun at the lower temperatures. This value is near the resolution for our instrument, and although there did seem to be a consistent trend in the EDX data, there is no detectable change in lattice constants. The sensitivity of the in-plane magnetic structure to synthesis conditions may account for some of the differences between previous Mossbauer and neutron measurements [9-18]. This sensitivity also suggests that the magnetic structure may have a large response to pressure or chemical doping.

Powder x-ray diffraction is performed using a PANalytical X'pert Pro diffractometer with Cu $K\alpha_1$ radiation equipped with an incident beam monochrometer (Cu $K\alpha_1$ radiation) and an Oxford PheniX closed-cycle helium cryostat. Elemental composition is measured with a Hitachi TM3000 scanning electron microscope with a Brucker Quantax 70 energy dispersive x-ray attachment. Neutron powder diffraction measurements are performed using the POWGEN time-of-flight diffractometer at the Spallation Neutron Source [22]. Powder samples are measured in vanadium sample cans at 100 and 300 K. Representation analysis is performed with the Sarah software suite [23] and the structure is refined with FullProf [24]. The magnetic structure is found to be the same colinear magnetic structure of ferromagnetic Kagome layers stacked antiferromagnetically along the c-axis as found previously by [16,18]. A refined moment of 1.85(2) $\mu_B$ per Fe is found at 100 K.

Magnetic data are collected from 1.8 to 750 K in applied magnetic fields from 0 - 60 kOe using a MPMS SQUID magnetometer from Quantum Design. Heat capacity, electrical and thermal transport measurements are made using a Physical Property Measurement System with various options, also from Quantum Design. Electrical contacts to the FeSn crystals are made with 50 $\mu$m diameter Pt wire using a silver epoxy (H20E) cured at 120 °C in air for 30 min. Contact resistances are between 1 and 5 ohms.

## III THEORETICAL METHODS

DFT calculations are performed using the projector augmented wave method [25] with the generalized gradient approximation in the parametrization of Perdew, Burke, and Enzerhof [26] for exchange-correlation as implemented in the Vienna ab initio simulation package (VASP) [27]. For Fe a standard potential is used (Fe in the VASP distribution), and for Sn a potential in which d states are treated as valence states, is used (Sn_d). In order to accommodate the layered AFM ordering as observed experimentally, we consider the experimental structure involving the doubled unit cell along the c direction. In most cases, we use an $8 \times 8 \times 6$ k-point grid and an E cutoff of 500 eV. The spin-orbit coupling, SOC, is included, but the +U correction is not included because FeSn is an itinerant magnetic system.

## IV. RESULTS AND DISCUSSION

As described above, the crystal structure of FeSn (Figure 1a,d) comprises Kagome nets of Fe stuffed with Sn and separated by Sn layers. The structure is hexagonal (space group P6/$mmm$) and the refined lattice constants from powder x-ray diffraction for crystals grown in this work are $a$ = 5.2959 Å, and $c$ = 4.4481 Å in good agreement with previous reports [7]. There was no detectable variation in the lattice constants for crystals quenched from temperatures between 600 °C and 720 °C. The temperature dependence of the lattice constants from 15-300 K are shown in Fig. 3. The much larger variation of the $a$ vs. c lattice constant is another indication of the 3D character of FeSn. In most true quasi-2D compounds, like graphite, the $c$ lattice constant is more sensitive to temperature [28]. The thermal expansion of FeSn at 280 K is $3.1 \times 10^{-6}$ K$^{-1}$ along the $c$ direction and $19 \times 10^{-6}$ K$^{-1}$ along the $a$ direction. The larger value is similar to that for soft metals like Sn or Al.

The refined powder neutron diffraction results from FeSn (Fig. 2) indicate an antiferromagnetic structure similar to that reported previously [16, 18] with Fe moments of 1.85(2) $\mu_B$ in the ab plane at 100 K. No canting of the moments out of plane was detected. From the powder diffraction data it was not possible to determine the in-plane direction of the ordered Fe moments.

Magnetization data from FeSn single crystals are summarized in Fig. 4 and Fig. 5. A maximum in the susceptibility and heat capacity [Fig. 6] at 365 K is taken as the Neel temperature. Well below $T_N$ the magnetic susceptibility is larger with H//c, which suggests that the ordered Fe magnetic moments mainly lay in the $a$-$b$ plane, consistent with previous descriptions of the magnetic structure [8-19] and our neutron data. With H in the $a$-$b$ plane the susceptibility below $T_N$ decreases with decreasing temperature, but only by about 30 % from $T_N$ to 2 K for crystals quenched from 600 °C. As noted in the methods section, for crystals quenched from 700 °C the susceptibility exhibits a much larger in-plane anisotropy resulting in nearly a 70% drop from $T_N$ to 2 K with H// (100) (not shown). This behavior is similar to that reported recently [19]. Some of the differences in the susceptibility for H in the a-b plane are likely due to different distributions of antiferromagnetic domains. The distribution of domains could change with synthesis conditions and certainly will change if the Fe moment direction changes with temperature from a high symmetry direction, such as the (100) [11], to a lower symmetry direction such as the (3.732, 1,0) [13].

The susceptibility above $T_N$ is isotropic (Fig. 4b) suggesting that orbital effects are small. The data are well-described by Curie-Weiss behavior [ C/(T-$\Theta_{CW}$)] with an effective moment of 3.4 $\mu_B$ per Fe, which is smaller than the effective moment expected for high spin $Fe^{2+}$ of 4.9 $\mu_B$. This is not surprising given FeSn is an excellent metal and one would expect some itinerant character to the magnetism. What is surprising is that $\Theta_{CW}$ = 173 K indicating substantial ferromagnetic interactions (also reported in [18]) in spite of the long-range antiferromagnetic order observed below 365 K. This likely reflects the much stronger in-plane ferromagnetic interactions. In addition, in the Fe-Sn system there are five known compounds :$FeSn_2$, FeSn, $Fe_3Sn_2$, $Fe_5Sn_3$ and $Fe_3Sn$. The compounds with Fe/Sn ratios greater than 1 are ferromagnetic, while compounds with Fe/Sn less than or equal to 1 are antiferromagnetic. This suggests competing ferromagnetic and antiferromagnetic interactions for FeSn [7, 29]. Magnetization data at 2 K and 300 K are shown in Fig. 5 for H//c and H//a. With H//c , M vs H is linear at all fields and temperatures, as is expected for a simple antiferromagnet. For H//a, however, there is a significant positive curvature at fields below 10 kOe and temperatures below $T_N$ suggesting a complicated magnetic structure in the basal plane. Although the curvature is less for FeSn crystals quenched from 600 °C as compared to crystals prepared at higher temperatures, it is still present.

The heat capacity data for FeSn are shown in Fig. 6 and exhibit a clear signature of a second order magnetic transition at $T_N$. The estimated entropy associated with $T_N$ is rather small, only about 17% of Rln2, suggesting that most of the entropy is removed above $T_N$ through short-range in-plane ferromagnetic order. The low temperature heat capacity indicates metallic behavior with an electronic contribution to the heat capacity of $\gamma T$ with $\gamma$ = 5.5 mJ/ $K^2$ mole atoms, similar to the values previously reported [19,30] and consistent with the expectations of theory, which predicts 3.5 mJ/$K^2$ mole atoms (see below). Assuming a simple Debye model and neglecting possible magnetic contributions to the heat capacity data below 10 K gives a Debye temperature of 311 K.

The resistivity with I//c and I//a is shown in Figure 7 for FeSn. Good metallic behavior is observed in both directions with residual resistivities of 1 $\mu\Omega$-cm and 1.6 $\mu\Omega$-cm and $\rho$(300 K)/$\rho$(2 K) ratios of 71 and 154 for I//c and I//a respectively. The excellent metallic behavior of FeSn has enabled recent de Hass van-Alphen (dHvA) investigations of the electronic structure [19, 20]. Only a small feature in the resistivity is observed at $T_N$. Although the crystal structure contains 2D Kagome layers [Fig. 1], $\rho_a/\rho_c \approx 3$ implying that the overall electronic structure is not quasi-2D and that the electronic conductivity along the $c$-axis via the Sn layers is significant. Resistivity data from the double Kagome layered compound $Fe_3Sn_2$ gives $\rho_a/\rho_c \approx 1$ [31], also indicating a relatively isotropic resistivity and electronic structure. Hall data are shown in Fig 8. The good metallic resistivity coupled with multiple bands near the Fermi energy make the Hall signal small and difficult to interpret. Within our experimental resolution the Hall signal is linear in H at all temperatures and is positive above 200 K, and negative below 200 K suggesting that both electron and holes bands contribute to the Hall signal. Part of the Hall signal could be due to an anomalous magnetic contribution, but the quality of our data precludes a more elaborate analysis. If it is assumed that at low temperatures a single band dominates the Hall signal, this yields a carrier density of 7.3 x $10^{21}$ electrons/$cm^3$. Multiple bands are also implied by the Seebeck data shown in Fig. 9 where the thermal gradient is along $c$ direction. The Seebeck data exhibit two sign changes as a function of temperature. The negative minimum near 30 K is likely due to phonon-drag, which typically occurs in this temperature range near a maximum in the thermal conductivity. Good conductors like Cu metal have small Seebeck features of similar magnitude in this temperature range[32]. The positive maximum in S near $\approx 175$ K is an indication of multiple bands [33, 34] as is the sign change in S near 350 K. Near $T_N$, [Fig 9b] the Seebeck data exhibit a small kink indicating a change in the electronic structure due to long-range antiferromagnetism. It is interesting that the Seebeck signal goes to zero fairly close to $T_N$ = 365 K. One would expect S=0 if the contribution from electrons and holes exactly cancelled, as it would in a perfectly compensated semimetal [33]. We also measured Seebeck data in an applied magnetic field of 8 T and found no change within experimental resolution.

The thermal conductivity from the same FeSn crystal is displayed in Figure 10. Near room temperature it is estimated assuming the Wiedemann-Franz (WF) relationship that roughly half of the heat is carried by phonons and half by electrons. Although the WF is not expected to be valid at intermediate temperatures, we show the data to make a point about thermal conductivity measurements. It is often assumed that in a metal the total thermal conductivity, $\kappa_{Total}$, can be separated into two contributions consisting of heat carried by electrons, $\kappa_e$, and heat carried by phonons, $\kappa_{ph}$, ie. $\kappa_{Total} = \kappa_{el} + \kappa_{ph}$. In systems with significant electron-phonon interactions, however, the separation is not possible [35]. Application of the WF relationship without considering this effect results in unphysical results such as a negative $\kappa_{ph}$ for temperatures between 10 K and 40 K, as shown in Fig. 10. This result suggest that electron-phonon coupling may be strong in FeSn.

To gain insight into the electronic and magnetic properties of FeSn, we carried out DFT calculations. We first examine 3 types of magnetic arrangements, layered-type AFM with spins lying in the crystallographic $a$ ($c$) direction [A-AFMa (A-AFMc)], FM with spins lying in the $a$ direction (FMa). It is found that A-AFMa has the lowest energy, in agreement with the experimental results, followed by A-AFMc and then FMa. This sequence of magnetic ground states implies that the energy scale of the out of plane AFM coupling is larger than the easy-plane single-ion anisotropy. The ordered Fe spin moment is found to be $\approx 2\mu_B$, corresponding to $S = 1$, and insensitive to the magnetic ordering. This value is consistent with the moment of $\approx 1.85(2)$ $\mu_B$ determined from our neutron diffraction measurement.

The band structure and the density of states (DOS) for the A-AFMa phase are presented in Fig. 11. Because of the SOC and out-of-plane AFM coupling, Dirac dispersions at the K points are gapped (massive) as indicated by the red circles. However, several Dirac dispersions at the H point remain massless (ungapped), as was also noticed in [21]. In particular, see red circle at H for energies about 0.3 eV below the Fermi energy. In addition to the Dirac-type dispersions, there appear rather flat dispersions along the $\Gamma$-M-K line at $\approx 0.5$ eV, $\approx -1.5$ eV, and $\approx -2.5$ eV from $E_F$ and along the A-L-H line at $\approx -1.5$ eV, $\approx -1.7$ eV, and $\approx -2.7$ eV from $E_F$ as indicated by blue rectangles. These are the characteristic features arising from an underlying Kagome lattice and consistent with very recent reports [20,21]. Along the layer-stacking $c$ direction, i.e., the $\Gamma$-A line, dispersive bands and flat bands coexist, and dispersive bands cross the Fermi level at non-zero in-plane momenta (not shown), supporting metallic conductance along the c direction. Reflecting the flat dispersions, the DOS consists of several peaks. One of peaks appears near the Fermi level and its height at the Fermi level is $\approx 17.9$ eV$^{-1}$(magnetic unit cell)$^{-1}$. This value corresponds to $\gamma \approx 3.5$ mJ K2/mole atoms, that is close to our experimental value. Comparing total DOS and partial DOS projected onto Fe d states, one notices that at $-4 \lesssim E - E_F \lesssim 2$ eV, the majority contribution is from Fe d states.

We now turn to the magnetic properties of FeSn. For this purpose, we consider an AFM state that is transferred from the A-AFMc by flipping a column of spins along the $c$ direction (C-AFMc) in addition to A-AFMa,c and FMa. The total energy of these magnetic orderings is mapped onto the following Heisenberg model:

$$H = J \sum_{\langle ij \rangle_{ab}} \mathbf{S}_i \cdot \mathbf{S}_j + J' \sum_{\langle ij \rangle_c} \mathbf{S}_i \cdot \mathbf{S}_j + K \sum_i |S_i^z|^2, \qquad (1)$$

where, $J$ ($J'$) is the in-plane (out-of-plane) exchange between nearest-neighbor Fe spins, and $K$ is the single-ion anisotropy. Using $S = 1$, we find $J \approx -41.2$ meV, $J' \approx 3.9$ meV, and $K \approx 0.03$ meV. While the electronic band structure suggests three-dimensional electronic character, the magnetic response is found to be strongly two-dimensional, $|J| \gg J'$. This could explain the positive Curie-Weiss temperature despite the A-AFMa ordering at low temperatures. Within a mean-field approximation with these $J$ and $J'$ and $S$=1 and neglecting the small $K$, $T_N$ is estimated to be $T_N = \frac{4}{3}(-2J + J') \approx 1335$ K, which is about three times higher than the actual $T_N$ = 365 K. This also indicates the importance of the strong two-dimensional spin fluctuations in supressing $T_N$.

Such a two-dimensional character is clearly seen in the magnon excitations. Fig. 12 shows the predicted magnon excitations within the nearest neighboring exchange model with the easy-plane anisotropy computed using the Holstein-Primakoff approximation [36] with the matching of matrix elements method [37,38] to treat the easy-plane anisotropy. As shown in Fig. 12, magnon dispersions are dominated by the in-plane FM coupling $J$, and the dispersion along the out-of-plane direction is very weak. As a consequence of the easy-plane anisotropy, the degeneracy is lifted at low-energy excitations at the $\Gamma$ point, leaving a gapless Goldstone mode corresponding to the in-plane spin precession and a gapped mode corresponding to the out-of-plane spin precession with the excitation energy $\approx 0.7$ meV. While the Goldstone mode could be gapped by the higher-order anisotropy that is supposed to be small, the splitting of the order of 1 meV could be resolved in future inelastic neutron scattering experiments.

## V CONCLUSIONS

This paper reports an experimental and theoretical investigation of the electronic, magnetic and thermodynamic properties of FeSn single crystals. From resistivity, Seebeck, Hall, thermal expansion measurements and theory we conclude the bulk electronic structure is 3D in spite of the layered appearance of the crystal structure. The electronic structure does contain Dirac nodes within $\approx 0.3$ eV below the Fermi energy. The Dirac dispersions at the K point in the Brillouin zone are massive (gapped), but the Dirac dispersions at the H point remain massless (ungapped). The magnetism, however, is highly 2D with $J_{in\text{-}plane} \approx -41.2$ meV and $J_{out\text{-}of\text{-}plane} \approx 3.9$ meV. Magnon dispersions are calculated, which gives rise to the interesting prediction of a gapless Goldstone magnon mode and a gapped magnon mode at the center of the Brillouin zone. These predictions can be tested using inelastic neutron scattering.

### Acknowledgements
This research was supported by the US Department of Energy, Office of Science, Basic Energy Sciences, Materials Sciences and Engineering Division. The neutron scattering measurements were conducted at the Spallation Neutron Source and were sponsored by the Scientific User Facilities Division, Office of Basic Energy Sciences, US Department of Energy. This research used resources of the Compute and Data Environment for Science (CADES) at the Oak Ridge National Laboratory, which is supported by the Office of Science of the U.S. Department of Energy under Contract No. DE-AC05-00OR22725.

References

1. J. X. Yin, S. T. S. Zhang, H. Li, K. Jiang, G. Q. Chang, B. J. Zhang, B. Lian, C. Xiang, I. Belopolski, H. Zheng, T. A. Cochran, S. Y. Xu, G. Bian, K. Liu, T. R. Chang, H. Lin, Z. Y. Lu, Z. Q. Wang, S. Jia, W. H. Wang, and M. Z. Hasan, *Nature* **562**, 91 (2018).
2. L. A. Fenner, A. A. Dee, and A. S. Wills, *J. Phys. Cond. Mat.* **21**, 452202 (2009).

3.  L. Ye, M. Kang, J. Liu, F. von Cube, C. R. Wicker, T. Suzuki, C. Jozwiak, A. Bostwick, E. Rotenberg, D. C. Bell, L. Fu, R. Comin, and J. G. Checkelsky, *Nature* **555**, 638 (2018).

4.  T. Kida, L. A. Fenner, A. A. Dee, I. Terasaki, M. Hagiwara, A. S. Wills, *J. Phys. Cond. Mat.* **23**, 112205 (2011).

5.  M. Yao, H. Lee, N. Xu, Y. Yang, J. Ma, O. V. Yazyev, Y. Xiong, M. Shi, G. Aeppli, and Y. Soh, arXiv:1810.01514.

6.  Z. P. Hou, W. J. Ren, B. Ding, G. Z. Xu, Y. Wang, B. Yang, Q. Zhang, Y. Zhang, E. K. Liu, F. Xu, W. H. Wang, G. H. Wu, X. X. Zhang, B. G. Shen, and Z. D. Zhang, *Adv. Mat.* **29**, 1701144 (2017).

7.  H. Giefers and M. Nicol, *J. Alloys and Compds.* **422**, 132 (2006).

8.  W. F. Ehret and A. F. Westgren, *J. Am. Chem. Soc.* **55**, 1339 (1933).

9.  C. Meyer, and K. Nagorny, *Ber. Bunsenges. Phys. Chem.* **93**, 1386 (1989).

10. H. Yamamoto, *J. Phys. Soc. Jpn.* **21**, 1058 (1966).

11. G. Trumpy, E. Both, C. Djega-Mariadassou, and P. Lecocq, *Phys. Rev. B.* **2**, 3477 (1970).

12. C. Meyer and K. Nagorny, *Hyper. Interact.* **47**, 457 (1981).

13. S. K. Kulshreshtha, and P. Raj, *J. Phys. Met. Phys.* **11**, 281 (1981).

14. L. Haggstrom, T. Ericsson, R. Wappling, and K. Chandra, *Phys. Scr.* **11**, 47 (1975).

15. S. Ligenza, *Phys. Status Solidi b* **50**, 379 (1972).

16. S. Ligenza, *Phys. Status Solidi b* **45**, 721 (1971).

17. S. Ligenza, *Phys. Status Solidi b* **44**, 775 (1971).

18. K. Yamaguchi and H. Watanabe, *J. Phys. Soc. Jpn.* **22**, 1210 (1967).

19. M. Kakihana, K. Nishimura, D. Aoki, A. Nakamura, M. Nakashima, Y. Amako, T. Takeuchi, T. Kida, T. Tahara, M. Hagiwara, H. Harima, M. Hedo, T. Nakama, and Y. Onuli, *J. Phys. Soc. Jpn.* **88**, 014705 (2019).

20. M. Kang, L. Ye, S. Fang, J. S. You, A. Levitan, M. Han, J. I. Facio, C. Jozwiak, A. Bostwick, E. Rotenberg, M. K. Chan, R. D. McDonald, D. Graf, K. Kaznatcheev, E. Vescovo, D. C. Bell, E. Kaxiras, J. van der Brink, M. Richter, M. P. Ghimire, J. G. Checkelsky, and R. Comin, arXiv 1906.02167.

21. Z. Lin, C. Wang, P. Wang, S. Yi, Q. Zhang, Y. Wang, Z. Wang, Y. Sun, Z. Sun, J.-H. Cho, C. Zheng, Z. Zhang, arXiv: 1906.05755

22. A. Huq, J. P. Hodges, L. Heroux, and O. Gourdon, Zeitschrift fur Kristallographie Proceedings **1**, 127 (2011).

23. A.S. Wills, Physica B 276, 680-681 (2000).

24. J. Rodriguez-Carvajal, Physica B 192, 55 (1993

25. G. Kresse and D. Joubert, Phys. Rev. B **59**, 1758 (1999).

26. P. Perdew, K. Burke, and M. Ernzerhof, Phys. Rev. Lett. **77**, 3865 (1996).

27. G. Kresse and J. Furthmüller, Phys. Rev. B **54**, 11169 (1996).

28. W. C. Morgan, *Carbon* **10**, 73 (1972)

29. B. C. Sales, B. Saparov, M. A. McGuire, D. J. Singh, and D. S. Parker, *Sci. Rep.* **4**. 7024 (2014).

30. M. Larsson, S. Bystrom, K. Marklund, and T. Lindqvist, *Physica Scripta* **9**, 51 (1974).

31. B. C. Sales, unpublished.

32. A. Hasegawa and T. Kasuya, *J. Phys. Soc. Jpn.* **28**, 75 (1970).
33. B. C. Sales, M. A. McGuire, A. S. Sefat, and D. Mandrus, *Physica C* **470**, 304 (2010)
34. H. J. Goldsmid, *Electronic Refrigeration* (Pion Limited, London 1986, pp. 29-48).
35. B. C. Sales, O. Delaire, M. A. McGuire, and A. F. May, *Phys. Rev.* **83**, 125209 (2011).
36. T. Holstein and H. Primakoff, Phys. Rev. **58**, 1098 (1940).
37. P.-A. Lindgård and O. Danielsen, J. Phys. C: Solid State Phys. **7**, 1523 (1974).
38. K. Tsuru, J. Phys. C: Solid State Phys. **19**, 016 (1986).

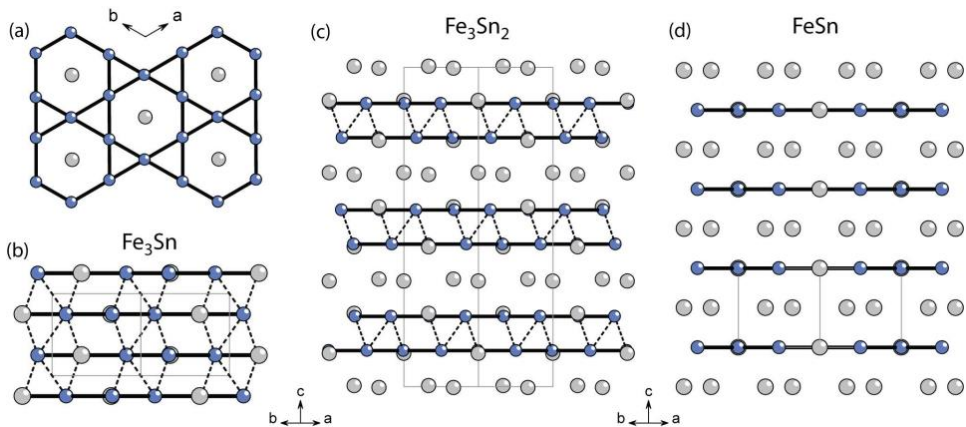

Figure 1. Kagome motifs in Fe-Sn compounds. (a) The Kagome net composed of corner shared triangles of Fe (blue) with hexagonal holes filled with Sn (gray) viewed normal to the 2D layer. The crystal structure of $Fe_3Sn$, $Fe_3Sn_2$, and FeSn are shown in (b), (c), and (d), respectively, with viewing direction within the Kagome planes. Heavy black lines denote intra-plane Fe-Fe bonds, while Fe-Fe connections between Kagome nets are shown as dotted lines. Bonds to Sn are omitted for clarity, and the gray lines in (b,c,d) outline the hexagonal unit cells.

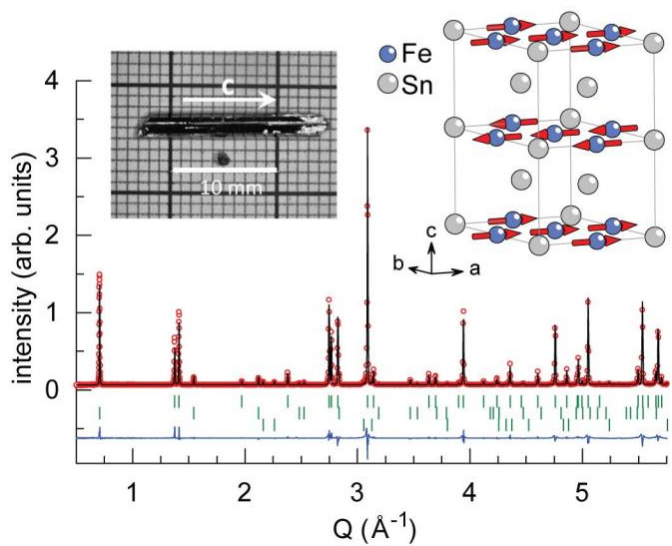

Figure 2. Refinement of powder neutron diffraction data from FeSn. From top to bottom the green ticks below the peaks correspond to the nuclear structure, the magnetic structure and a 5% Sn impurity. The refined moment per Fe at 100 K is 1.85(2) $\mu_B$. The antiferromagnetic structure is shown in the inset and corresponds to Fe moments aligned ferromagnetically in each plane, although the in-plane direction of the spins could not be determined from the refinement. Examples of FeSn crystals investigated are shown in the left inset.

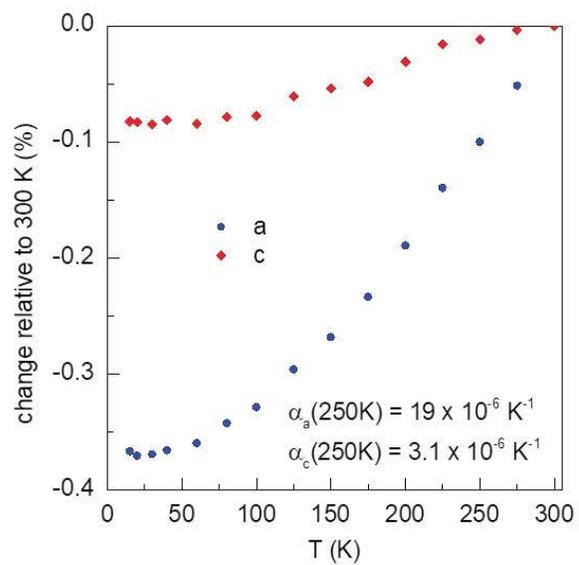

Figure 3. Temperature dependence of the FeSn lattice parameters.

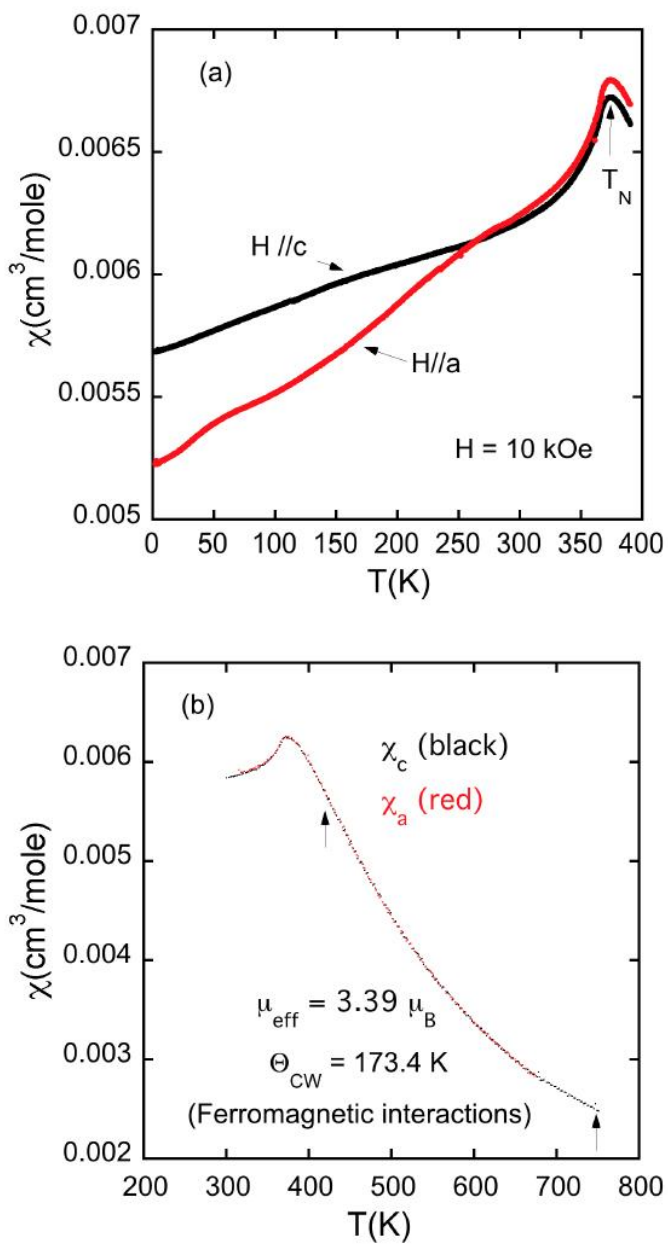

Figure 4. (a) Magnetic susceptibility vs. temperature of FeSn crystals quenched from 600 °C with the magnetic field directed along the *c* or *a* axis for temperatures between 2 and 400 K. (b) Magnetic susceptibility vs. temperature of FeSn crystals from 300 to 750 K. The data between the black arrows is fit to a Curie-Weiss law.

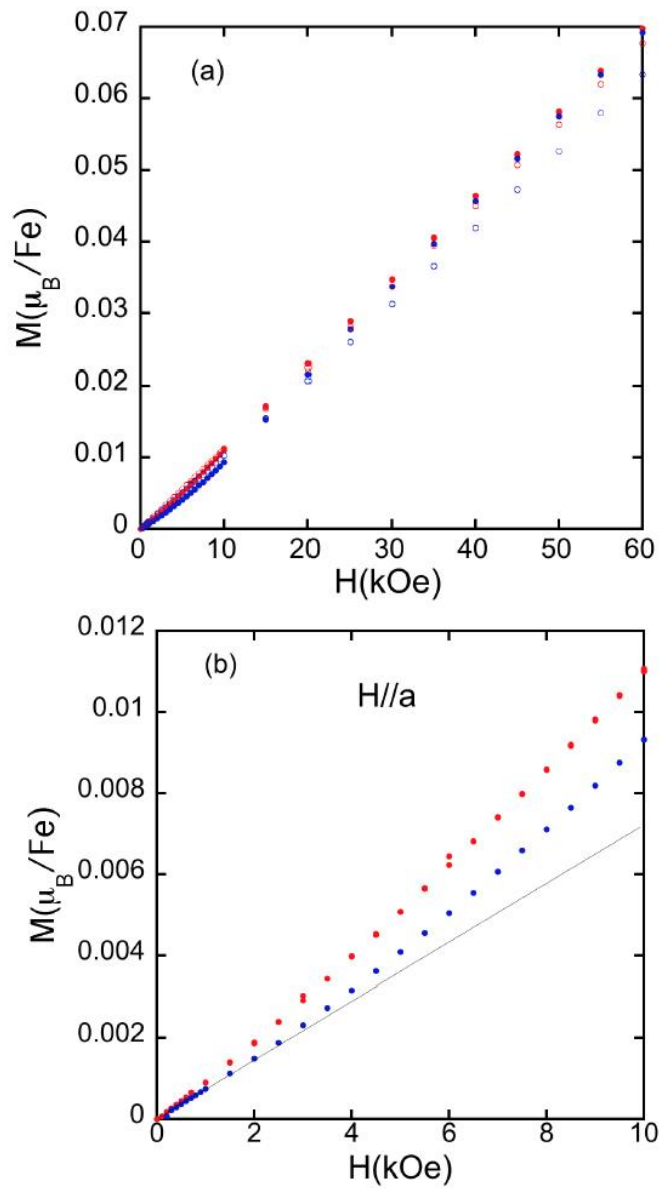

Figure 5. Magnetization vs magnetic field (a) Red 300 K, Blue 2 K, solid symbols H//a, open symbols H//c (b) Low field behavior for H//a. There is a significant upward curvature for magnetization curves with H//a. The black line is a linear fit to the 2K data for applied magnetic fields less than 1 kOe.

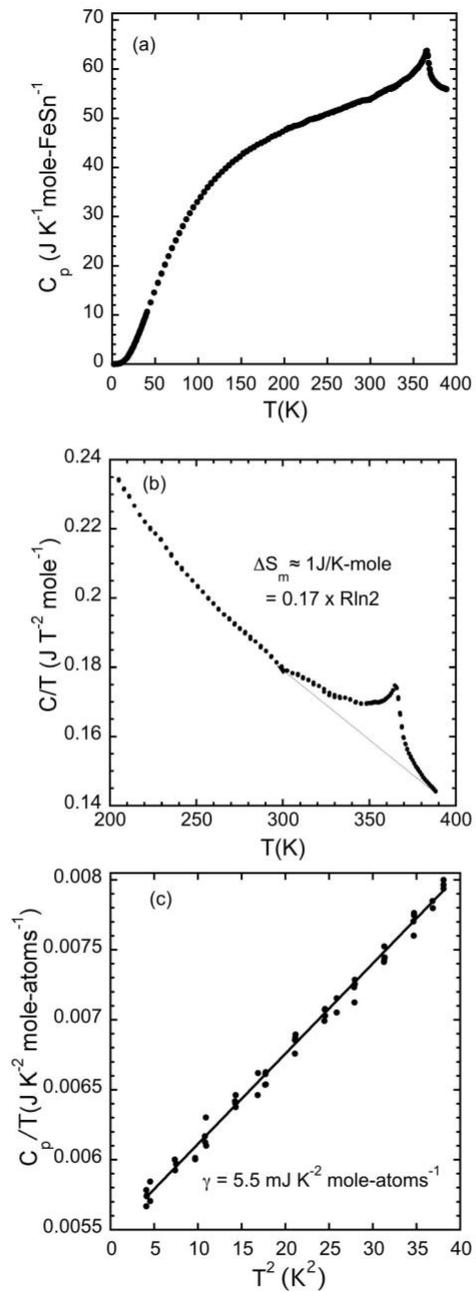

Fig. 6 Heat capacity data from FeSn. (a) $C_p$ vs T for temperatures between 2 and 380 K (b) Rough estimate of magnetic entropy involved in magnetic transition near 365 K. (c) Cp/T vs $T^2$ for data below 6.5 K

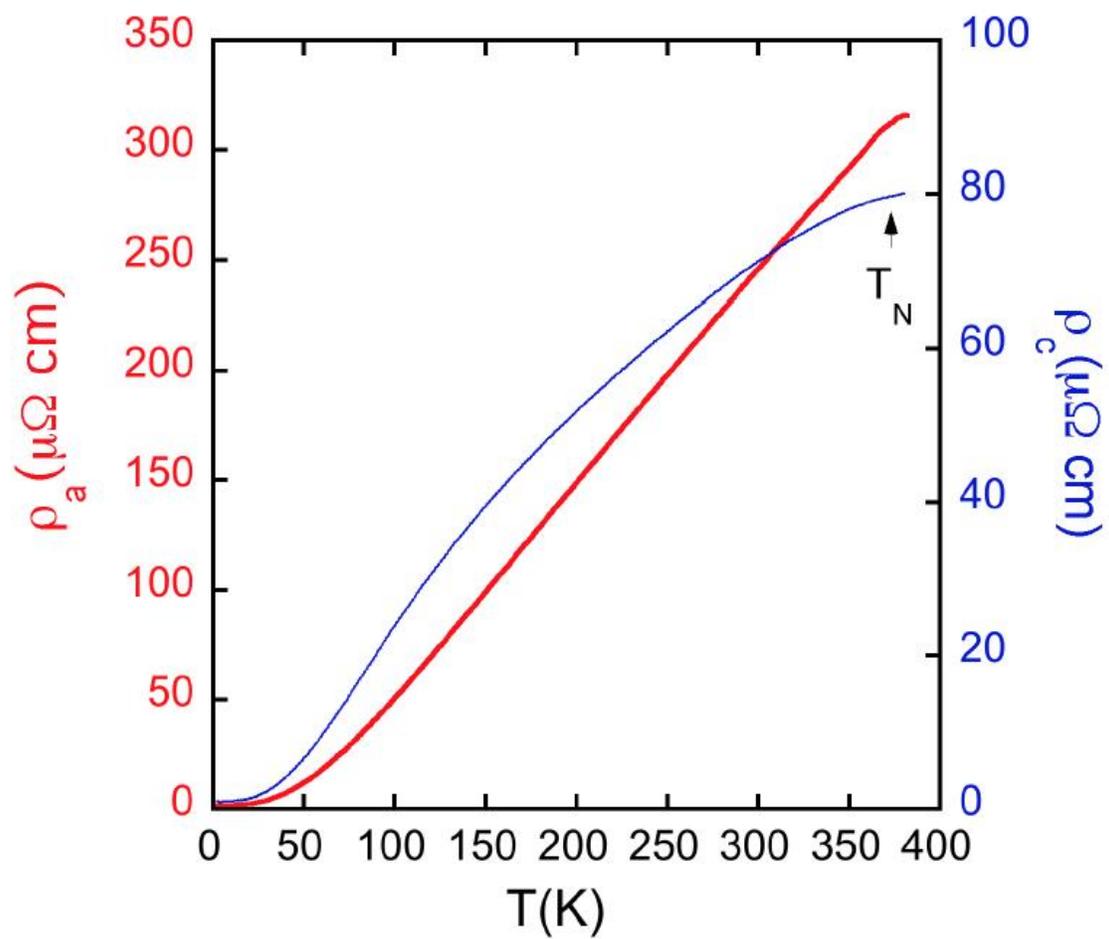

Fig. 7. Resistivity vs. temperature of FeSn with I//a and I//c.

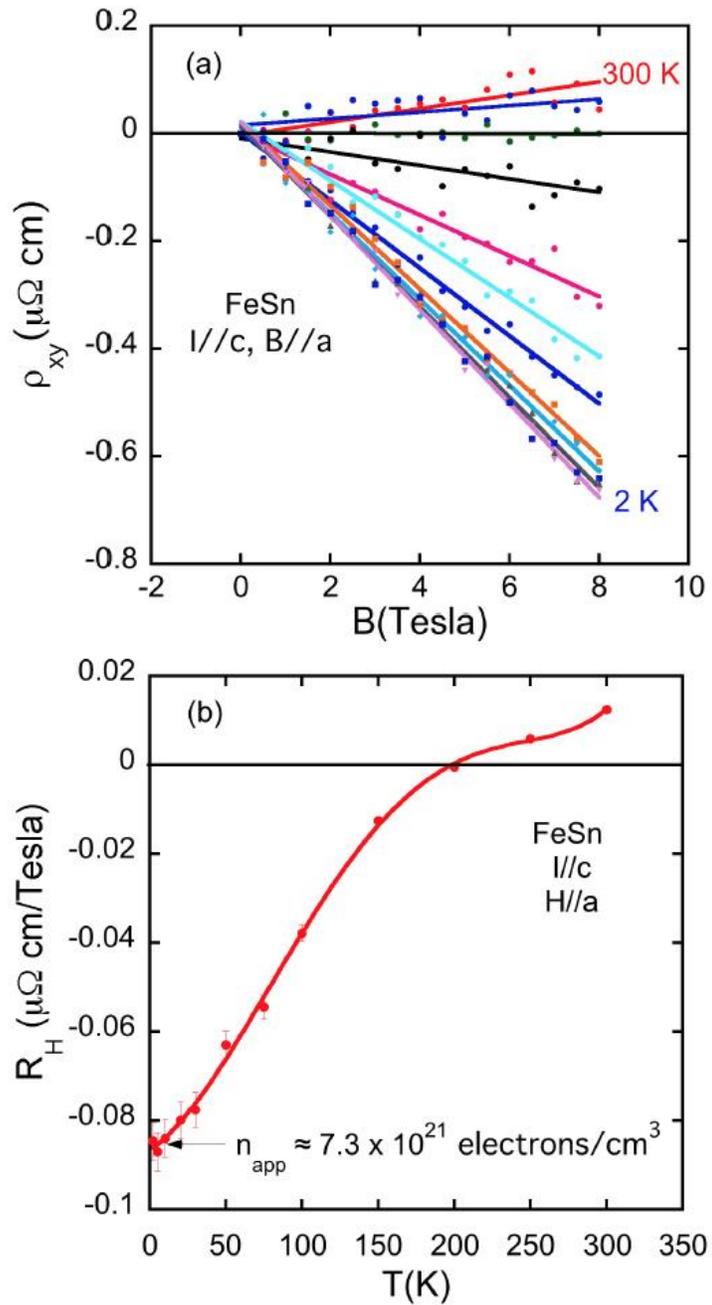

Figure 8. (a) Hall resistivity of FeSn versus applied magnetic field for temperatures between 2 and 300 K, with I//c and B//a.  (b) Temperature dependence of Hall coefficient as determined from data in (a).

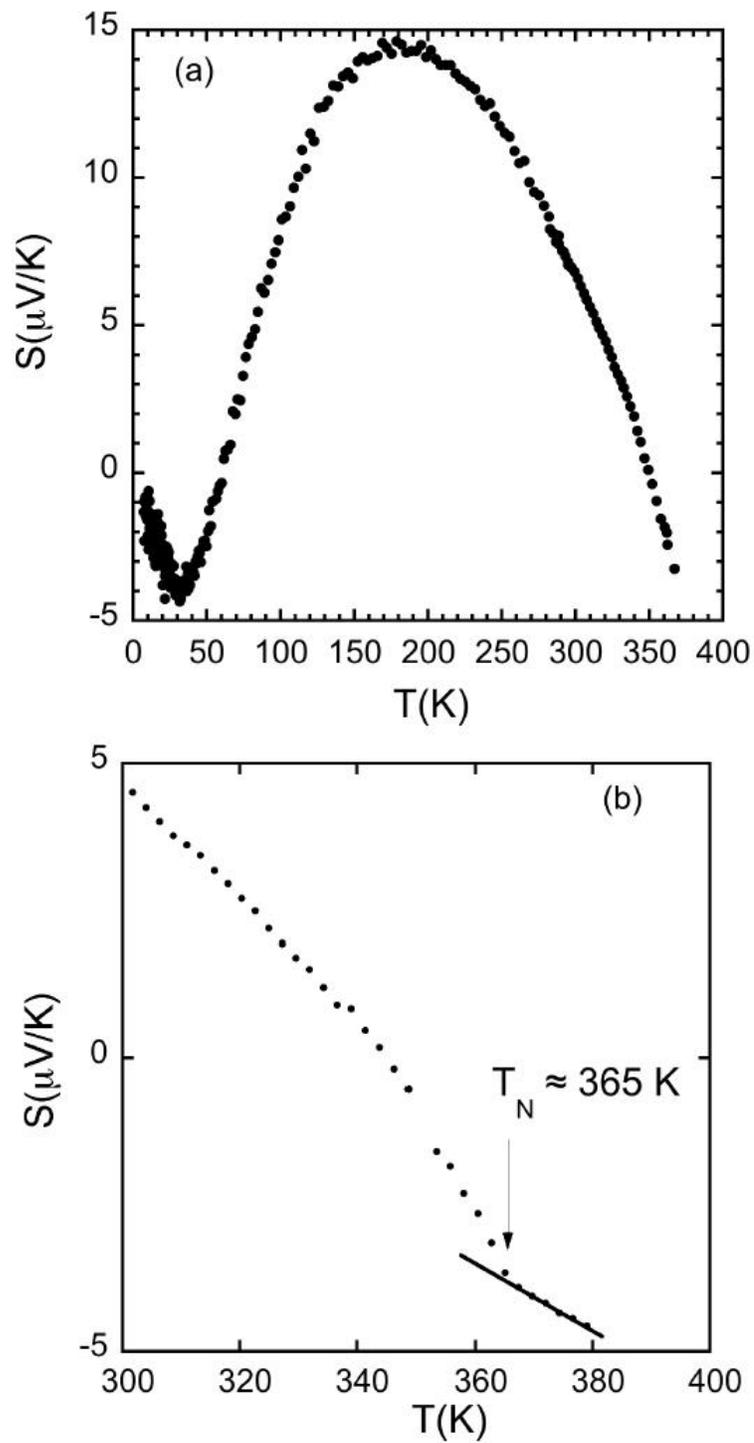

Figure 9 (a) Seebeck coefficient of FeSn crystal vs. temperature from 2 to 370 K. (b) Seebeck coefficient versus temperature near the Neel temperature.

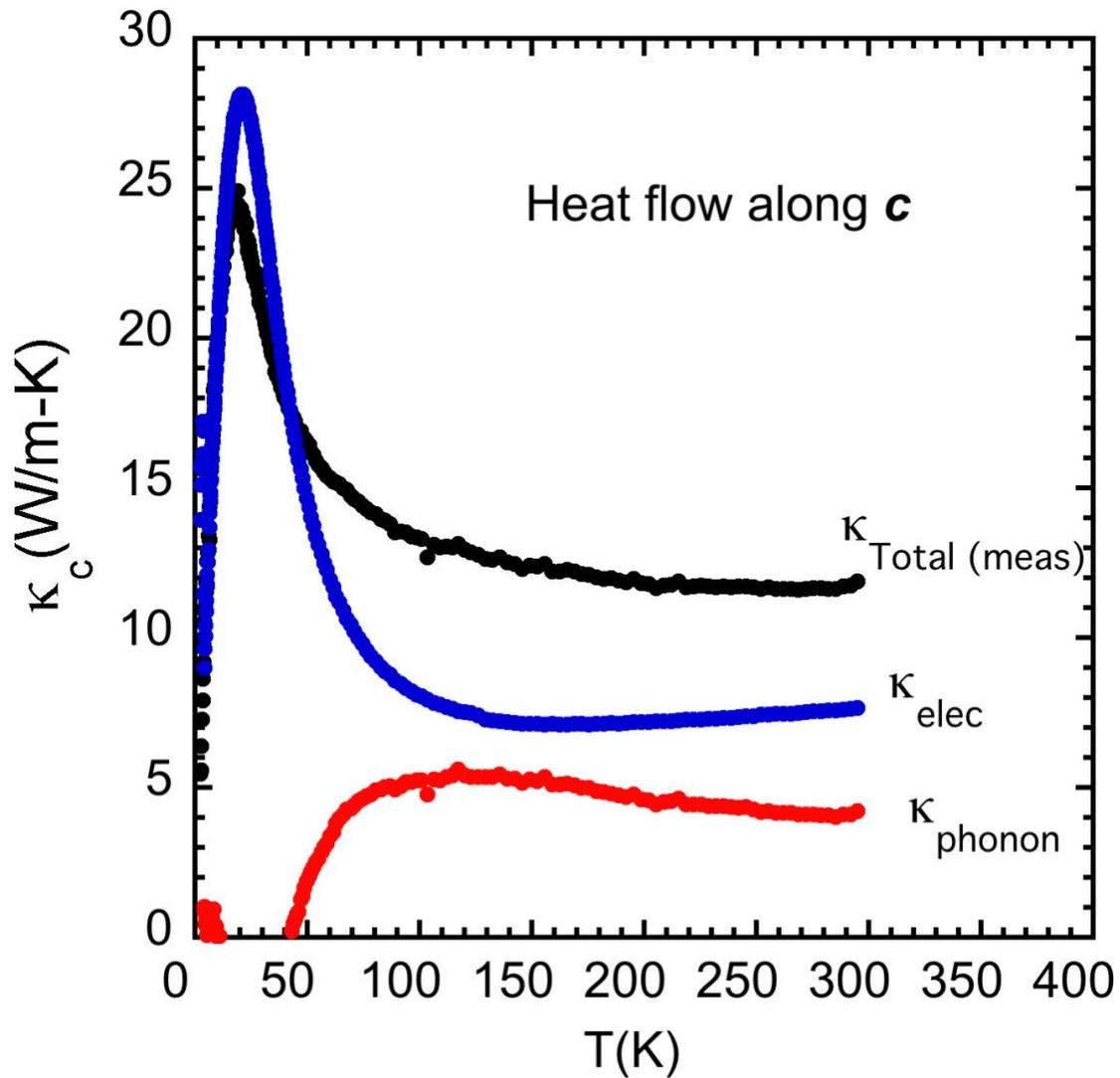

Figure 10. Thermal conductivity of a FeSn crystal from 2 to 300 K with heat flow along the *c* direction. Estimate of the electronic and lattice contributions to the thermal conductivity if the WF relationship is used. See text for more details.

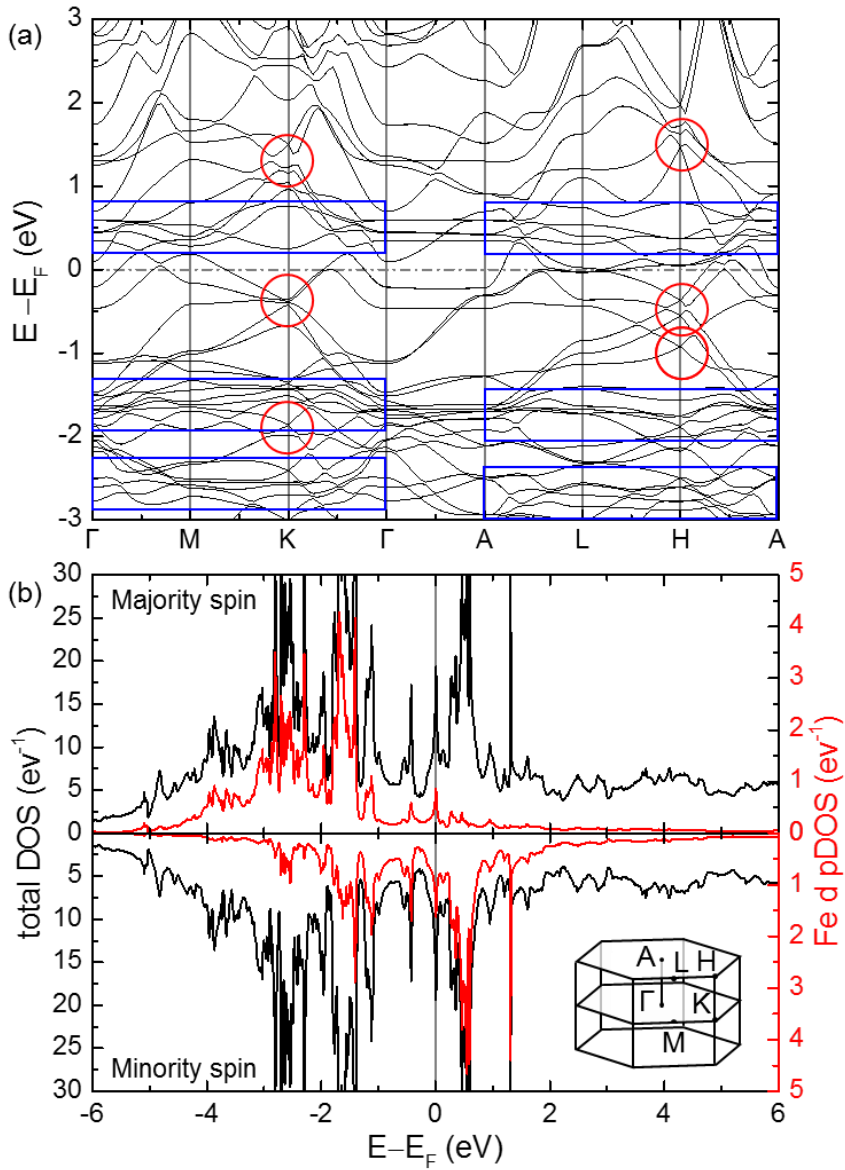

Figure 11. DFT results for FeSn in the AAFMa phase, (a) band structure and (b) total DOS and partial DOS projected on Fe d states. In (a), gapped Dirac dispersions (K point) and ungapped Dirac dispersions (H point) and flat bands arising from a kagomé lattice are indicated by red circles and blue rectangles, respectively. The inset of (b) shows the magnetic Brillouin zone. Details of the DFT calculations are given in the main text.

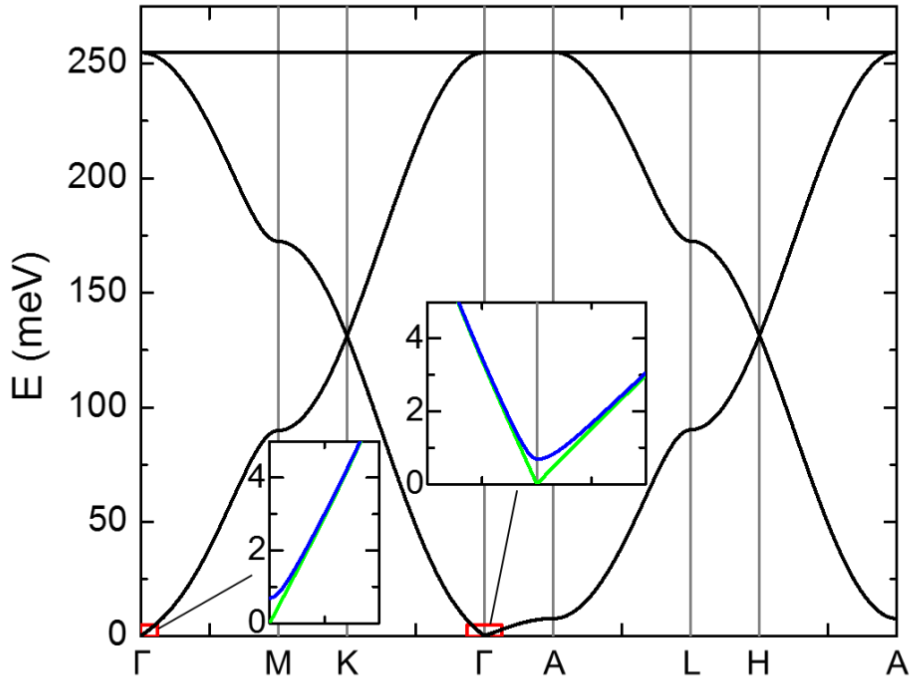

Figure 12. Predicted magnon dispersions from the effective spin Hamiltonian in Eq. (1) with the in-plane (out-of-plane) exchange coupling $J = -41.2$ meV ($J' = 3.9$ meV), the easy-plane anisotropy $K = 0.03$ meV, with $S = 1$. Due to the easy-plane anisotropy, the degeneracy is lifted near the $\Gamma$ point, where the in-plane mode remains gapless (green line in the insets) while the out-of-plane mode is gapped (blue line).